# The driving force behind genomic diversity

Salla Jaakkola[1], Sedeer El-Showk[1], Arto Annila[1,2,3,]*

[1]*Department of Biosciences,* [2]*Institute of Biotechnology and* [3]*Department of Physics, FI-00014 University of Helsinki, Finland*

**Abstract**

Eukaryote genomes contain excessively introns, inter-genic and other non-genic sequences that appear to have no vital functional role or phenotype manifestation. Their existence, a long-standing puzzle, is viewed from the principle of increasing entropy. According to thermodynamics of open systems, genomes evolve toward diversity by various mechanisms that increase, decrease and distribute genomic material in response to thermodynamic driving forces. Evolution results in an excessive genome, a high-entropy ecosystem of its own, where copious non-coding segments associate with low-level functions and conserved sequences code coordinated activities. The rate of entropy increase, equivalent to the rate of free energy decrease, is identified with the universal fitness criterion of natural selection that governs populations of genomic entities as well as other species.

The discovery of mobile genetic elements by McClintock in the 1940s and introns by Sharp and Roberts in 1977 challenged the once predominant view of a genome as a plain repository of biological information [1,2,3]. Since then, many mechanisms have been found - particularly in eukaryotes - which are capable of increasing, decreasing and redistributing genomic material [4] beyond simple insertion and deletion; examples include gene duplication, transfer of genetic material, polyploidy, genesis of genes [5], exon shuffling [6], intron gain and loss [7,8]. Despite increasing understanding of evolutionary mechanisms that shape the genome, the vast amount of non-coding sequences such as B-chromosomes, pseudogenes, transposons, short repeats, introns and miscellaneous unique sequences, remains perplexing. Also it is puzzling, why the size of genome does not correlate with the complexity of an organism [4].

The selfish DNA theory takes a bold stance by picturing *all* sequences as replicating entities in mutual competition for survival [9,10,11]. The view of a genome as an ecosystem of its own is insightful and consistent with the theory of evolution by natural selection [12]. Obviously the genome is open to external influence, *e.g.*, affecting allele frequencies but the genome-centric view, despite considering externalities only implicitly, provides understanding to the evolution of a genome toward diversity.

In this study we consider the possibility that genomes are driven to the diversity of sequences in the quest of increasing entropy. The general thermodynamic principle underlies many spontaneous phenomena that are referred to as *natural processes* [13]. Since no system, irrespective of its evolutionary mechanisms, can escape the 2$^{nd}$ law of thermodynamics, also processes in a genome should be described as diminishing potential energy differences, *i.e.*, as consuming free energy in interactions. This is the essence of theory of evolution by natural selection [12] that was recently formulated in thermodynamic terms [14] to account for diverse natural phenomena and puzzle of nature [15,16,17]. We consider the imperative of increasing entropy as a sufficient reason to explain *why* genomes organize into nested hierarchies of diverse sequences and display skewed distributions of coding and non-coding sequences.

## 1. Genome as a thermodynamic system

The 2$^{nd}$ law of thermodynamics merely states that potential energy differences tend to vanish in mutual interactions. Increase in entropy means dispersal of energy, not univocally increasing disorder as is often erroneously assumed. The principle of increasing entropy makes no difference between abiotic and biotic, although we tend to label as living those systems that attain and maintain high-entropy non-equilibrium states by coupling to external energy. The external energy provides the potential gradient that is consumed in raising the concentrations of complex entities, such as genes, beyond those at equilibrium. The complex just as simple entities are mechanisms that diminish the potential energy differences in interactions. They exist due to their functional properties that contribute to the consumption of free energy in the quest for stationary state in their surroundings.

Now that the 2$^{nd}$ law of thermodynamics has been formulated as an equation of motion [14], an evolutionary course, such as growth of a genome, can be understood and simulated. The evolution of a genome can be regarded as an energy-powered dissipative motion via chemical reactions. The seemingly dull quest for increased entropy is in fact a highly functional criterion. It selects from diverse energy transduction mechanisms those that will consume free energy most rapidly. Genes associate with powerful energy transduction mechanisms via expression, but also all other genomic sequences consume free energy, *e.g.*, in replication. The rate of entropy increase is regarded as the universal fitness criterion of natural selection that governs also populations of genomic entities.

The direction of genomic evolution, just as other evolutionary processes, toward more probable distributions can be deduced from the logarithmic probability measure known as entropy (see ref. 14 for derivation)

$$S = R\ln P = R\sum_{j=1}\ln P_j \approx \frac{1}{T}\sum_{j=1} N_j \left(\sum_k \mu_k - \mu_j + \Delta Q_{jk} + RT\right) \quad (1)$$

by comparing various distributions of genomic entities $j$, *i.e.*, sequences, in multiple copies $N_j$. Each genomic entity associates with chemical potential [18] $\mu_j = RT\ln[N_j\exp(G_j/RT)]$, where the Gibbs free energy $G_j$ is relative to average energy $RT$ per mole. For example, a conserved sequence associates with a high $G_j$ value which is particularly evident when a change in the sequence, *e.g.*, a mutation, collapses the entire energy transduction of the organism. In contrast, a nucleotide change in a miscellaneous sequence does not couple markedly with the overall energy transduction hence its associated $G_j$ is low. Clearly, it would take a great amount of information to obtain all $\mu_j$ values for an organism by including all terms in the summation of Eq. 1. However, even without precise knowledge of what specific genomic entities might be present at a given time and how they may propagate, it is possible to deduce the direction of evolution and ensuing overall distribution of the genomic entities by requiring that $S$ will increase until $dS/dt = 0$.

Genomic entities are transformed by reactions from one class to another. For example, a gene may mutate to a pseudogene but the most apparent flows of matter and energy happen during replication when nucleotides polymerize to sequences using external energy. Substrates, indexed by $k$ according to stoichiometry in Eq. 1, yield the product $j$ as long as the potential energy difference experienced by entity $j$, known as affinity $A_j = \sum\mu_k + \Delta Q_{jk} - \mu_j > 0$. Also the external energy $\Delta Q_{jk}$ is a substrate that couples to chemical reactions that transform sequences $N_k$ to sequences $N_j$. Typically the energy influx to the genome appears as high substrate potentials $\mu_k$, *i.e.*, matter $N_k$ of internal energy $G_k$, associated with triphosphates whose breakdown drives various chemical reactions.

According to Eq. 1, genomic material will accumulate as a result of numerous chemical reactions as long as there are supplies for it, *i.e.*, free energy. The growth toward a non-equilibrium stationary state will level off when potential energy differences vanish with respect to the surroundings. Conversely, the genome will begin to degrade toward the equilibrium when the coupling to external energy is broken.

The concept of average energy $RT$ is not limited to equilibrium systems, but can be computed for any ensemble that is sufficiently static [19]. Likewise, the second law of thermodynamics is not limited to isolated systems that, in fact, cannot dissipate and change partitions, and thus, cannot evolve. This is in contrast to the misconception that entropy would be a valid concept only for a closed system. The reason for the misunderstanding is that the earlier derivations of $S$ have aimed to deduce only the equilibrium partition where the free energy terms have vanished. Therefore the driving forces of evolution and its directional nature have remained obscure.

## 2. An evolving genome

Statistically energy can only be dispersed down along gradients. Entropy increases also in a genome when energy is distributed among diverse genomic entities and surroundings by various reactions. For example, the genome is primarily coupled to sources of free energy via genes. Contemporary genomes draw matter and energy by indirect actions of numerous gene products, *i.e.*, proteins that facilitate diverse flows of energy. Presumably, the primordial energy transduction mechanisms were simpler and directly involved nucleic acids in dissipative processes as has been articulated in the RNA world hypothesis [20]. However, chemical syntheses are not non-directional as often mistaken but take the direction of increasing entropy. Therefore an evolutionary course to an integrated energy transduction system, where nucleic acids primarily carry information and proteins are mostly responsible for dissipation, need not to be improbable when the surroundings are high in energy and abundant with ingredients. Despite the directional driving force, the appearance of organisms, *i.e.* intricate energy transduction machinery, may take a long time. While mechanisms of dissipation have evolved over eons, the thermodynamic imperative for the simple as well as for the complex system has remained the same. Energy flows by diverse mechanisms to the genome or from the genome depending on thermodynamic gradients (Fig. 1).

A genomic entity $j$, *e.g.*, a gene, regulatory element, transposon, intron, codon, *etc.*, contributes to the overall rate of evolution toward more probable states by facilitating flows $v_j = dN_j/dt$. The master equation of evolution [14]

$$\frac{dS}{dt} = \sum_j \frac{dS}{dN_j}\frac{dN_j}{dt}$$
$$= \frac{1}{T}\sum_{j=1}\frac{dN_j}{dt}\left(\sum_k \mu_k - \mu_j + \Delta Q_{jk}\right) = \frac{1}{T}\sum_{j=1} v_j A_j \quad (2)$$

describes flows $v_j$ as responses to the driving forces $A_j$. Each genomic entity $j$ contributes by the rate $(dS/dN_j)(dN_j/dt)$ to the growth of entropy. According to the fitness criterion of natural selection, given by Eq. 2, evolution channels via mechanisms which contribute most to $S$ [14]. This characteristic process is pictured within the selfish DNA theory and the theory of evolution by natural selection in general as the mutual competition for survival. Accordingly, genomic sequences that are able to access free energy resources by their characteristic mechanisms will survive. Thus the genome is similar to any other ecosystem where interdependent entities assemble from and disassemble to common constituents depending on free energy that they access.

The flows of matter to the genome and from it

$$v_j = r_j \frac{A_j}{RT} \quad (3)$$

are proportional to the free energy $A_j$ to satisfy the continuity equation [14]. The non-linear form of free energy gives rise to non-linear flows. When $A_j > 0$ ($A_j < 0$), also $v_j > 0$ ($v_j < 0$). Near the stationary state, *i.e.* the dynamic stationary state $A_j \approx 0$ and also $v_j \approx 0$. The rate coefficient $r_j$ depends on the mechanism of energy transduction, *e.g.*, catalysis. Some genes associate with powerful mechanisms, *e.g.*, enzymes whereas some non-coding sequences, such as short interspersed nuclear elements, link to replication mechanisms of others [21]. Miscellaneous sequences are mostly devoid of much means to conduct energy. For all mechanism $r_j > 0$ because every mechanism can be regarded as a result of an earlier natural process. For example, an enzyme results from a folding process that is a natural process too [17].

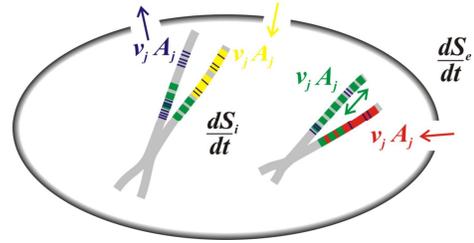

**Fig. 1.** Genome as an open system. Each genomic entity $j$ (colored) in numbers $N_j$ contributes to entropy by the flow rate $v_j$ and free energy $A_j$. As long as the free energy $A_j$ remains positive, matter flows in and entropy increases more rapidly within the genome than in its surroundings, $dS_i/dt > dS_e/dt$; Thus the genome continues to grow in size. When the gradients reverse, the genome will begin to loose its entities.

During the course of evolution, driving forces will vary and mechanisms will appear and disappear, affecting the flows of matter to and from the genome. In accordance with LeChâtelier's principle [18], when external conditions change, the system will take a new course toward a new attractor, which is the current most probable distribution of genomic entities. When $v_j$ of Eq. 3 is inserted to Eq. 2 and because $r_j > 0$ the 2$^{nd}$ law is indeed found $dS/dt = R\sum r_j(A_j/RT)^2 \geq 0$. There is no need to explain the rise of orderly structures by invoking an exemption that entropy would decrease in a living system at the expense of its surroundings. Entropy is increasing in living systems as well by dispersal of energy.

Using the master equation of evolution (Eq. 2) we may only outline the evolution of a genome in the statistical sense, not in mechanistic details. This inability does not only stem from a lack of knowledge about the fine details of the system, but from the intrinsic non-integrability of interdependent natural processes. A specific trajectory, *i.e.*, an evolutionary course, cannot be known in details. A small random variation at an early time may redirect the long-term course. When $dS/dt > 0$, the genome will grow; however, we cannot predict when an organism happens to acquire new genomic material. Evolution of an open system is non-deterministic and chaotic by its nature [14].

The thermodynamic description of a genome may at first appear naïve as if overlooking biological mechanisms. However, Eq. 1 is extremely detailed by denoting constituents of the system by every quantum of energy, in forms of matter and fields as well as all interactions. Mechanisms have no inherent value by the principle of increasing entropy. They are only means to devour free energy and to move toward more probable states. If there is a thermodynamic force and a mechanism to consume it, the system will evolve, irrespective of how the motive force is generated and how the mechanism of motion is implemented. The only relevant thermodynamic property of a genomic entity is its contribution to this energy transduction (Fig. 1). This raison d'être is consistent with the selfish gene perspective. According to the principle of sufficient reason there need not to be other incentives but to level differences in energy one way or another to explain the rise of genomic diversity.

### 3. Simulated genomic evolution

We simulated evolution of a genome that initially housed only one short, five residue, sequence $N_5 = 1$ associated with Gibbs free energy $G_5$. The exterior to the genome was modeled to comprise of base constituents in numbers $N_1$ and Gibbs free energy $G_1$ as well as external energy that may couple to the reactions. The starting point is not particularly important or crucial but it can be regarded as a model of primordial conditions where a short sequence happened to assemble due to a random synthesis. In our model system any two sequences $i$ and $k$ may assemble to any longer sequence $j$, and any sequence $j$ may disassemble to any two shorter sequences $i$ and $k$ randomly depending on the sign of $A_j$.

The endoergic syntheses of sequences were programmed according to Eq. 3, so that an assembly step of $j$ from substrates $k$ consumed an external quantum $\Delta Q_{jk}$. The formation of an entity $j$ was catalyzed by other entities $n \geq j$ so that the rate coefficient $r_j \propto \Sigma \mu_n \exp(-\mu_n)$ was proportional to $\mu_n$ weighted by its thermodynamic partition. This way each mechanism of energy transduction itself was modeled as a skewed distribution resulting from an earlier natural process. Our choice for the specific form of $r_j$ is unimportant but consistent. In the statistical sense a long and conserved but not necessarily continuous sequence associates with more powerful mechanisms to increase entropy than a short and non-conserved one. This is expressed in the form of $r_j$.

Fragmentation and breakdown of sequences were modeled as spontaneous random exoergic reactions. Any sequence could break apart at any point, *i.e.*, the probability distribution was uniform. The precise knowledge of mechanisms that shape the genome is not important to outline the overall course of evolution since the rate criterion of Eq. 2 will ensure that all mechanisms contributing to $S$ will be naturally selected during the course of evolution.

At each time step chemical potentials were calculated from the $N_j$ and $G_j$ values. Then a next step of aforementioned syntheses and degradations took place according to Eq. 3. The chemical potentials and free energy terms $A_j$ were updated for the following step. During the course of evolution the probabilities $P_j$ (Eq. 1) kept changing because syntheses and degradations of interdependent genomic entities were coupled. The 'memory' of the past course was contained in the energy reservoirs and their differences directed the future course. It was not modeled in, *e.g.*, as in a Markovian process with additional parameters. Entropy, the total amount of matter in the genome and its partitions among diverse entities were monitored but not used in any way to direct the course.

During the course of simulated evolution (Fig, 2) the initial sequence fragmented into shorter sequences that continued to grow, assemble and fragment anew. As long as the overall free energy remained positive the genome grew in size. Matter and energy flowed in and organized to genomic entities. The cumulative curve of genomic matter rose initially nearly exponentially because for a small system the supplies of free energy appeared almost unlimited in relation to its mechanistic capacity to raise its chemical potential by synthesizing new sequences. Subsequently, when more and more powerful mechanisms emerged via syntheses, the growth followed an approximate power-law form (straight line in a log-log plot) and finally turned to a logistic curve when the supplies narrowed and it became increasingly difficult to draw more matter to the system. For the non-integrable growth curve there is no general analytic form but the logistic curve, despite being deterministic, is a good approximation [15].

Eventually when all potential energy gradients were exhausted, the maximum entropy state, the maturity [22] was reached. The resulting genome housed copious low-energy sequences (small $G_j$) with low-level functions, (small $r_j$) and highly functional, (high $r_j$) and high-energy, (large $G_j$) sequences. The thermodynamically costly conserved sequences remained in the genome because without them the genome could not exist for very long at all. Also non-conserved sequences remained in the genome although they contributed only little - but did not cost much either. In other words, the non-equilibrium steady-state distribution reflected the thermodynamic balance among all entities by a skewed nearly log-normal distribution that has been described earlier [15].

The values of parameters $N$, $G_j$ and $\Delta Q_{jk}$ did affect outcomes of simulations. For example, a small total amount of matter ($N=\Sigma jN_j$), little external energy ($\Delta Q_{jk}$) and high Gibbs free energies ($G_j$) gave rise to a small genome with a narrow diversity, whereas large supplies of matter and energy as well as low-cost syntheses gave rise to a large genome with a large diversity of entities. However, irrespective of the parameter values, all systems evolved to diversity along fast routes of entropy increase by available energy transduction mechanisms and finally emerged with the maximum entropy distribution with the skew characteristic of natural distributions [15].

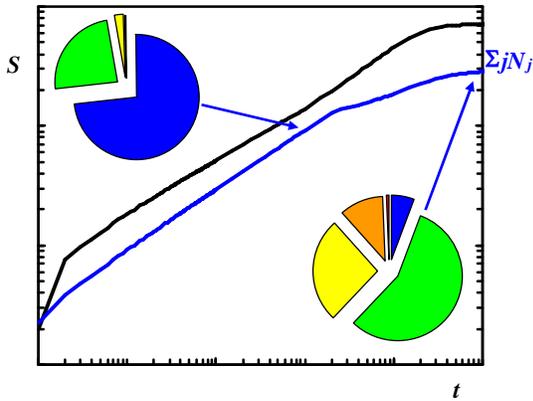

**Fig. 2.** Simulated evolution of a genome, its total size $\Sigma jN_j$ and entropy $S$ vs. time $t$ on the log-log scale. During the evolution the total amount of matter in base entities distributed between the genome and its surroundings according to the rates of entropy increase. The original genome was modeled to house only one short sequence. Initially entropy increased when new sequences emerged from the existing sequences via diverse mechanisms. The genome grew in size $\Sigma jN_j$ as long as free energy allowed, *i.e.*, $dS_i/dt > dS_e/dt$ (Fig. 1). Color-coded pie-charts illustrate abundance of diverse genomic entities in the middle and at the end of the simulation. Very long conserved sequences are colored in red, increasingly shorter sequences are coded by orange, yellow and green, and very short sequences in blue.

A specific natural system evolves using its particular set of mechanisms for genomic intake, outflow and rearrangements. These details of energy transduction vary from one organism to another but the principle is the same. Irrespective of mechanisms, energy will flow down along steep gradients. This imperative alone will result in the characteristically skewed distribution of genomic entities in agreement with data [23,24].

When external conditions change, free energy may reverse so that the prior non-equilibrium steady state becomes improbable, impossible to maintain. Subsequently the genome as an open system will shift its course toward new states, by discarding matter to match the decreased external supply. This process is customarily referred to as adaptation. The genome, jus as any other ecosystem, will diminish the reversed free energy by downsizing. Both, the coding and non-coding sequences as well as associated mechanisms of energy transduction are affected when the evolution turns its course toward a new non-equilibrium state. Changes in external conditions may be brought about by other open systems, *i.e.*, organisms that have acquired more efficient energy transduction mechanisms to draw from the common pool of resources.

## 4. Distribution of genomic entities

We find it reasonable to assume that most contemporary genomes are evolving slowly and thus display quasi-stationary distributions of entities

$$dS \approx 0 \Leftrightarrow \sum_k \mu_k - \mu_j \approx 0 \Leftrightarrow N_j \approx \prod_k N_k \, e^{(-\Delta G_{jk} + \Delta Q_{jk})/RT}. \quad (4)$$

that are examined by genome sequencing projects. In the same manner, we extracted distributions of genomic entities from the simulations and found them skewed resembling log-normal distributions (Fig. 3). The underlying principle is universal; natural processes lead to natural distributions that peak at the entity classes that are most efficient in leveling differences in energy. Chemical energy distributes among chemical entities via chemical reactions just as kinetic energy distributes among gas molecules via collisions.

When there are many mechanisms to reach a stationary state, the distribution of genomic entities is nearly continuous. When a particular system has only few efficient mechanisms to increase its entropy, the resulting distribution is sparse. The overall skewed form of a natural distribution is independent of the organism-specific mechanistic details. In eukaryotes low-functionality elements, miscellaneous sequences and remnants of transposons make the lower fraction of the skewed distribution; transposons and other replicating elements make the abundant middle fraction; and highly functional elements, the genes are in the small high fraction [4,25]. In prokaryotes low-functionality non-coding elements are almost entirely absent because these non-genic sequences would, without a nuclear compartment, severely compromise the vital energy transduction by high-functionality genes. Nevertheless, the characteristic skewed form is the same as exemplified by the distribution of *E. coli* genes [23] in accordance with the simulated distribution. Also consistently with our reasoning that the conserved sequence length correlates with the entropy increase functionality, the conserved, most functional sequences peak at higher lengths than non-conserved sequences [24].

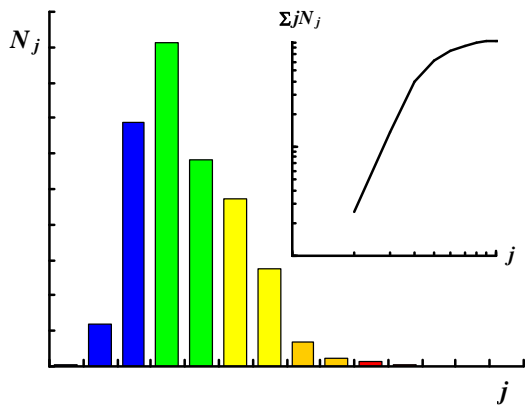

**Fig. 3.** Distribution of genomic entities for a model system at the end of simulated evolution. The high-$j$ end houses highly functional entities that increase entropy by recruiting matter and energy to the genome. They are thermodynamically expensive to maintain, hence low in numbers $N_j$. The natural distribution peaks at less conserved and shorter entities that generate $dS/dt$ by redistributing matter in the genome. The low-$j$ end contains entities with only little functionalities to increase entropy. The inset displays the cumulative curve of matter in genomic entities during evolution. The approximate power-law to logistic characteristic is common to natural processes [15].

## 5. Genome as an ecosystem

Even thought the main objective of this study was to show that thermodynamics alone gives rise to the genomic diversity, it was of course tempting to relate the above described thermodynamic classification based on the rate of entropy increase with the established classification of genomic entities [21,26]. However, the classification by the rates of entropy increase does not correspond one-to-one with the classification schemes of genes and non-genic elements e.g. based on various mechanisms of replication. The thermodynamic classification evaluates only the rates of entropy increase irrespective of how the energy transduction is accomplished. Keeping this in mind, we suggest the high tail of skewed distribution corresponds to eukaryotic genes and sets of networked genes as well as their associated highly conserved control elements. This fraction contains information to assemble most efficient machineries, i.e. organisms. These entities contribute to entropy by generating most influxes to the genome to support it. They associate with very efficient mechanisms (high $r_j$) but are low in numbers ($N_j$) because they are thermodynamically speaking costly (large $G_j$) to make and maintain. These costs amount, *e.g.*, during replication that is equipped with correction and proof reading mechanisms to ensure the vital functions of an organism.

We remind that the thermodynamic classification does not recognize entities by their mechanisms only by their contribution to *S*. Therefore, also long transposons with highly conserved and functional terminal regions are expected to be in the middle upper part of the distribution just as are less conserved genic regions. The most copious comparatively short sequences we associate with numerous transposons of various kinds. They contribute to entropy mainly by redistributing matter within the genome. These numerous intra-genomic operations ensure that no potential differences will develop within the genome, for example, that genes will not lump into only one chromosome. The very short sequences in the lower part of the distribution we relate to stretches of bases in miscellaneous unique sequences, *e.g.*, introns, intergenic segments and remnants of transposons. They are almost devoid of any functions to redistribute or recruit matter and energy. They contribute to the system's entropy by their numbers; otherwise, matter essentially flows out of the genome via them.

It is no new thought, but still insightful, to regard a genome as an ecosystem of its own as is done in the selfish DNA theory. According to thermodynamics, a comparison of a genome, *e.g.*, to a forest is not an allegory but a mere transformation in scale. At all levels of natural hierarchy the imperative to diminish free energy is the same. Genes are like trees that are responsible for most of ecosystem's energy transduction. We are hardly surprised to find diverse insects feeding on trees just as we should not be surprised to find, *e.g.*, transposons thriving on genomic material furnished by genes. Obviously insects are not all 'harmful' to the trees but *e.g.*, pollinate them, just as transposons are not merely devouring genetic material but may also bring about genomic rearrangements that give rise to new genes that may provide access to more free energy. All sources of free energy and all types of mechanisms qualify to fuel the rise of diversity.

## 6. Discussion

The principle of increasing entropy by decreasing free energy expressed by the thermodynamics of open systems is a powerful imperative to understand evolution of complex systems. Genomic evolution displays the common attributes of natural processes, most notably the sigmoidal courses and characteristically skewed distribution of genomic entities. The thermodynamic description of an evolving genome as an open system does not question the cumulating contemporary knowledge about the intricacies of genomic machinery. It reveals the mere consequences of the 2$^{nd}$ law of thermodynamics.

Thermodynamics of open systems provides the physical rationale for the selfish DNA theory, the insightful view of evolving genomes. However, the selfish DNA is not regarded as a parasite or as an end in itself [10]. Both non-coding and coding sequences are mechanisms of energy transduction that emerge from the natural process due to the same imperative. A fragmented genome with non-coding sequences incorporates more matter and energy into the system increasing its entropy. In this respect, the evolution of nuclear compartment [27] appears critical for enabling continued growth of the genome by including thermodynamically 'cheap' non-genic elements. The excessive genome provides possibilities for novel interactions that may result in new mechanisms by which to increase the system's entropy further. Thus diversity builds on diversity.

The results derived from thermodynamics are consistent with the findings that non-coding sequences, sometimes broadly referred to as junk-DNA, are not simply random non-sense sequences but exhibit a functional hierarchy, albeit not as sophisticated as that of translated sequences. The most elementary non-coding sequences hold only little functionality to increase entropy, whereas the some sophisticated sequences, *e.g.*, transposons, associate with a wealth of energy transduction machinery to consume free energy. Despite not being translated and also when not realized in phenotypes, these sequences are subject to natural selection that weeds on the basis of entropy increase rate [14]. Although the evolutionary pressure, *i.e.*, free energy coupling to eukaryote non-coding entities may not be as stringent - or more precisely, the gradient in the entropy landscape may not be as steep - for coding sequences, the principle is the same. Thermodynamics finds no demarcation line between the coding and non-coding sequences.

According to thermodynamics the amount of non-coding sequences may vary substantially from species to species [28,29,30] as long as diverse non-genic sequences do not interfere much with means of energy transduction due to gene expression. Entropy may increase over eons by proliferating and diversifying non-coding sequences as long as the organism has mechanism to access and devour free energy. In other words, the organism is just 'fit' to maintain and grow an extensive genome. It is cheap to expand with non-coding sequences. On the other hand, expansion by genes would require higher sources of free energy and greater returns in energy transduction in order to maintain them in the genome.

A long successful evolutionary history of ancient and still flourishing eukaryotes has accumulated high-entropy genomes. In certain cases, the entropic drive may have found no better alternative but to swell the genome. This offers an explanation to why the genome size does not correlate with the complexity of an organism. The entropy principle as a rationale for this "C-value enigma" clarifies why genomes tend to grow but not how this may happen. Mechanisms are not obtained from thermodynamics but once some have emerged, *e.g.*, due to random variation, they are valued by their rate of entropy increase. For example, when a chromosome multiplication takes place by either by polyploidy or by cellular division, relative frequencies of the two mechanisms depend on their relative rates of entropy increase. We expect exhaustive eukaryotic genome inventories to provide data to critically and quantitatively assess the entropy interpretation of genome evolution by revealing more detailed distributions of all genomic entities.

Intuitively we tend to see more sense in the small number of highly functional coding sequences and less so in numerous low-activity non-coding sequences. However, the line of demarcation vanishes when we recognize the universal motif of activities, to level differences in energy. The presented thermodynamic view of an evolving genome, just as any other dissipative system, as energy transduction machinery is a change in the paradigm [31,32] from monistic biology to holistic natural sciences that moves from genetic determinism to non-deterministic evolution governed by the 2$^{nd}$ law of thermodynamics. Free energy powers many motions as was foreseen by Boltzmann [33].

**Acknowledgment**

We thank Mahesh Karnani, François Sabot, Ilari Sundberg and Peter Würtz for valuable comments and corrections.